\documentclass[%
 reprint,
superscriptaddress,
 amsmath,amssymb,
 aip, apl
]{revtex4-2}

\usepackage{graphicx}
\usepackage{dcolumn}
\usepackage{bm}

\usepackage{makecell}
\usepackage{capt-of}

\makeatletter
\def\maketitle{
\@author@finish
\title@column\titleblock@produce
\suppressfloats[t]}
\makeatother

\makeatletter
\def\maketitleSupp{
\@author@finish
\title@column\titleblock@produce 
}
\makeatother

\begin{document}

\preprint{APS/123-QED}

\title{Unidirectional propagation of zero-momentum magnons}

\author{Ond\v{r}ej Wojewoda}
\email{ondrej.wojewoda@vutbr.cz}
\affiliation{CEITEC BUT, Brno University of Technology, Purky\v{n}ova 123, Brno, 612 00, Czech Republic\\}%
\author{Jakub Holobr\'{a}dek}
\affiliation{CEITEC BUT, Brno University of Technology, Purky\v{n}ova 123, Brno, 612 00, Czech Republic\\}%
\author{Dominik Pavelka}
\affiliation{Faculty of Mechanical Engineering, Institute of Physical Engineering, Brno University of Technology, Technick\'{a} 2, Brno, 616 69, Czech Republic}
\author{Ekaterina Pribytova}
\affiliation{Faculty of Mechanical Engineering, Institute of Physical Engineering, Brno University of Technology, Technick\'{a} 2, Brno, 616 69, Czech Republic}
\author{Jakub Krčma}
\affiliation{Faculty of Mechanical Engineering, Institute of Physical Engineering, Brno University of Technology, Technick\'{a} 2, Brno, 616 69, Czech Republic}
\author{Jan Kl\'{i}ma}
\affiliation{Faculty of Mechanical Engineering, Institute of Physical Engineering, Brno University of Technology, Technick\'{a} 2, Brno, 616 69, Czech Republic}
\author{Jan Michali\v{c}ka}
\affiliation{CEITEC BUT, Brno University of Technology, Purky\v{n}ova 123, Brno, 612 00, Czech Republic\\}
\author{Tom\'{a}\v{s} Lednick\'{y}}
\affiliation{CEITEC BUT, Brno University of Technology, Purky\v{n}ova 123, Brno, 612 00, Czech Republic\\}
\author{Andrii V. Chumak}
\affiliation{Faculty of Physics, University of Vienna, A-1090 Wien, Austria}
\author{Michal Urb\'{a}nek}
\email{michal.urbanek@ceitec.vutbr.cz}
\affiliation{CEITEC BUT, Brno University of Technology, Purky\v{n}ova 123, Brno, 612 00, Czech Republic\\}%
\affiliation{Faculty of Mechanical Engineering, Institute of Physical Engineering, Brno University of Technology, Technick\'{a} 2, Brno, 616 69, Czech Republic}
\date{\today}

\begin{abstract}
We report on experimental observation of unidirectional propagation of zero-momentum magnons in synthetic antiferromagnet consisting of strained CoFeB/Ru/CoFeB trilayer. Inherent non-reciprocity of spin waves in synthetic antiferromagnets with uniaxial anisotropy results in smooth and monotonous dispersion relation around Gamma point, where the direction of the phase velocity is reversed, while the group velocity direction is conserved. The experimental observation of this phenomenon by intensity-, phase-, and time-resolved Brillouin light scattering microscopy is corroborated by analytical models and micromagnetic simulations. 

\end{abstract}

\maketitle


Synthetic antiferromagnets (SAFs) recently attracted a lot of attention in the magnonics community, as they can provide additional degrees of freedom for spin wave dispersion engineering compared to conventional ferromagnetic systems \cite{ishibashi2020switchable, shiota2020tunable, mouhoub2023exchange, chumak2022advances, pirro2021advances, gerevenkov2023nonreciprocal, wintz2016magnetic, Verba2019}. One of the most prominent features of dispersion relation in SAFs is non-reciprocity caused by the dipolar field interaction between the layers \cite{grunberg1981magnetostatic, grunberg1986layered, cochran1990brillouin, vavassori2000brillouin, millo2023unidirectionality, chen2019excitation, gallardo2021spin, Gallardo2024}. Spin wave non-reciprocity and symmetry breaking in material systems are of fundamental interest and can also be exploited in practical magnonic applications, such as diodes or circulators \cite{Devolder2023}. Non-reciprocal dispersion can be achieved also in ferromagnets, e.g., by exploiting interfacial effects \cite{zakeri2010asymmetric, di2015direct, vanatka2021spin, gladii2016frequency}, or current induced Doppler shift \cite{vlaminck2008current, kim2021current}. However, the non-reciprocity in SAFs is much more pronounced. Another way to create non-reciprocal spin-wave system is to use a grating coupler \cite{chen2019excitation, temdie2023high}, but, these systems can be tuned only to a specific $k$-vector and frequency. 
Nevertheless, the methodology for modeling spin waves in SAF's is not fully developed and not all consequences of nonreciprocal dispersion relation are explored.

\begin{figure}
\includegraphics{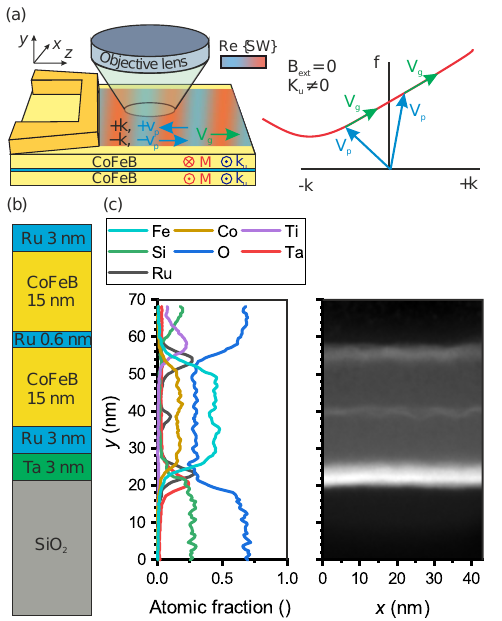}
\caption{\label{fig:Sketch} Experimental geometry. (a) Schematics of the experiment and unidirectional excitation of spin waves with positive and negative wavevectors. Due to the linear dispersion, the group velocity direction is not changed upon the reversal of wavevector direction. (b) Schematics of the studied system. The two CoFeB layers are coupled antiferromagnetically through RKKY interaction. (c) STEM image of the lamella prepared under the antenna layer (right panel), and atomic fraction of individual elements obtained by EDX analysis.}
\end{figure}

In this paper, we report on the observation of smooth and monotonous spin wave dispersion in the vicinity of $\Gamma$ point ($k$=0) in SAFs. This property leads to unidirectional spin wave propagation over a wide range of frequencies. The group velocity points in the same direction at both sides of the $\Gamma$ point, whereas the phase velocity is reversed upon the $\Gamma$ point crossing (see Fig.~\ref{fig:Sketch}a), resulting in a situation where we can observe unidirectional energy flow with two opposing phase velocities (wavevectors). The smooth and monotonous dispersion causes that even zero-momentum spin waves (with $k$=0) propagate with nonzero group velocity. This phenomenon exhibits itself as a change of the sign of phase evolution for the same direction of the spin wave propagation, and is directly observed by phase-resolved Brillouin light scattering (BLS) microscopy \cite{Vogt2009}.

To study this behavior we prepared a sample consisting of a~multilayer stack of two 15\,nm thick CoFeB layers separated by a 0.6\,nm thick Ru spacer layer which were grown by magnetron-sputtering on Si(100) substrate covered by 285\,nm thick thermal SiO$_2$ layer (Fig.~\ref{fig:Sketch}b). On top of the SAF layer, we patterned a 500\,nm wide microwave antenna for spin wave excitation using electron beam lithography and lift-off process. The antenna consists of a stack of Ti 5\,nm/SiO$_2$ 10\,nm/Ti 5\,nm/Au 50\,nm thin films deposited by e-beam evaporation. The SAF layer was imaged by scanning transmission electron microscopy with a high-angle annular dark-field detector (STEM-HAADF) showing individual layers with Z-contrast and its chemical composition was investigated by energy-dispersive X-ray spectroscopy (STEM-EDX), see Fig.~\ref{fig:Sketch}c. The STEM image shows good quality of the interfaces and measured distances of 15\,nm
between Ru interlayers revealed by the EDX chemical composition profile well correspond with the nominal thicknesses of the layers [$t_{\mathrm{CoFeB}}=(15\pm1)\,$nm]. 

\begin{figure}
\includegraphics{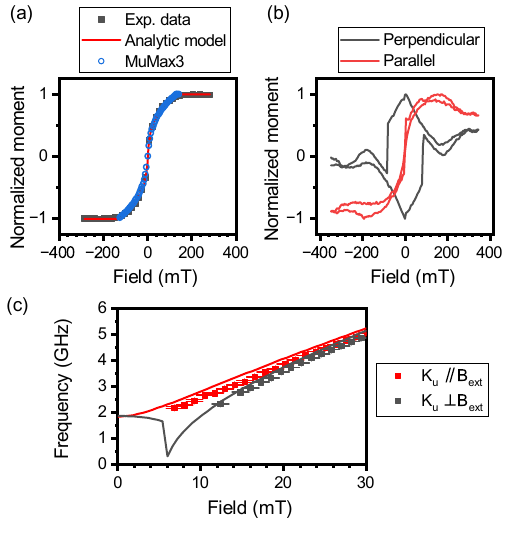}
\caption{\label{fig:StaticChar} Static characterization of SAF sample. (a) In-plane VSM hysteresis curve, analytic modeling based on energy minimalization, and MuMax3 simulation results. (b) Hysteresis loops measured by wide-field Kerr microscope. The external field was applied perpendicular to uniaxial anisotropy axis. The black (red) curve shows magnetization perpendicular (parallel) to the external field. (c) Position of the out-of-plane polarized (acoustic) mode measured with the external field parallel and perpendicular to the uniaxial anisotropy axis and relevant analytical modeling. }
\end{figure}

In-plane (IP) and out-of-plane (OOP) hysteresis loops of the sample were acquired by vibrating sample magnetometry (VSM), see Fig.~\ref{fig:StaticChar}a and supplementary Fig.~S1. To obtain static magnetic parameters of the SAF layer, we model the hysteresis loop using minimization of the energy of two macrospins in a similar manner as it is done in Stoner-Wohlfarth model \cite{wohlfarth1948mechanism, skomski2008simple}. The total energy reads as \cite{gallardo2019reconfigurable, githubSWT}
\begin{equation}
E=E_\mathrm{dip}+E_\mathrm{Zeeman}+E_\mathrm{ex}+E_\mathrm{anis}+E_\mathrm{J}+E_\mathrm{inter},
\label{eq:TotEnergy}
\end{equation}
where $E_\mathrm{dip}$ is dipolar energy, $E_\mathrm{Zeeman}$ is energy caused by external field, $E_\mathrm{ex}$ is intralayer exchange, $E_\mathrm{anis}$ is energy of uniaxial anisotropy, $E_\mathrm{inter}$ is dipolar energy between the two layers, and interlayer exchange $E_\mathrm{J}=J_\mathrm{bl} (1-m_1\cdot m_2 )+J_\mathrm{bq}[1-(m_1\cdot m_2 )]^2$ , where $J_\mathrm{bl}$ and $J_\mathrm{bq}$ are bilinear and biquadratic exchange constants, respectively \cite{slonczewski1991fluctuation, guslienko1994bilinear}. Using this model, we obtained saturation magnetization ($M_\mathrm{s}$) and interlayer coupling constants from the simultaneous fitting of the experimental IP and OOP hysteresis loops. The thickness of the layers was estimated from TEM measurements, and anisotropy was estimated from offset of out-of-plane polarized (acoustic) mode resonance. The fitting results are summarized in Table~\ref{tab:Param}. The observed uniaxial anisotropy arises probably from the Ta seed layer roughness \cite{cui2013tuning, Scheibler2023}.

\def\@fnsymbol#1{\ensuremath{\ifcase#1\or *\or \dagger\or \ddagger\ord
   \mathsection\or \mathparagraph\or \|\or **\or \dagger\dagger
   \or \ddagger\ddagger \else\@ctrerr\fi}}
\newcommand{\ssymbol}[1]{^{\@fnsymbol{#1}}}


We also implemented bilinear and biquadratic coupling into MuMax3 code \cite{Vansteenkiste2014design} using custom fields. For this, we have to derive the effective field caused by the coupling and switch off the standard exchange interaction between the layers. The discretized effective coupling field $B_\mathrm{J}$ then reads as
\begin{equation}
B_\mathrm{J} = \frac{\partial E_\mathrm{J}}{\partial \boldsymbol{M}}=\frac{ \sum_i \boldsymbol{m}_i - \boldsymbol{m}}{c_z M_\mathrm{s}} \left( J_\mathrm{bl} + 2 J_\mathrm{bq} \sum_i \boldsymbol{m_i} \cdot \boldsymbol{m} \right),
\end{equation}
where $c_z$ is cell size in the OOP direction, and $i$ is the layer index. The results (see Fig.~\ref{fig:StaticChar}a, circles) are in perfect agreement with the experiment and results from Eq.~\ref{eq:TotEnergy}, which proves that the implementation of the coupling constants in MuMax3 is valid, and the code can also be used for simulations of dispersion relations.

To get further insight into the sample static behavior, we performed wide-field Kerr microscopy with the external magnetic field applied at different angles to the easy axis of the SAF. The magnetization components perpendicular and parallel to the applied field direction are shown in supplementary Fig.~S2, together with corresponding loops obtained from the energy minimalization. When the external magnetic field is applied perpendicularly to the easy axis, it is possible to stabilize the single domain state, with the Neél vector aligned with the easy axis, see Fig.~\ref{fig:StaticChar}b. This property of the studied system is essential for later BLS experiments which are performed in zero external magnetic field. 

Angle-resolved Kerr microscopy experiments revealed that the switching is deterministic and depends on the magnetic field sweep direction (see supplementary Fig.~S3). This deterministic switching is unexpected in perfectly symmetric SAFs. Here, it is probably caused by slightly different magnetic properties of the individual layers or interfaces \cite{jang2010magnetic}, and thus slightly different total energies for the two opposite directions of magnetization. It could be also caused by different layer thicknesses; however our STEM data (Fig.~\ref{fig:Sketch}c) do not show any difference above the measurement uncertainty of $\pm1\,$nm. 

\begin{table*}
    \centering
    \setlength\tabcolsep{0pt}
    \caption{Fitted and fixed material parameters. $\mathrm{t}_\mathrm{CoFeB}$ is thickness of the individual CoFeB layers, $\mathrm{t}_\mathrm{Ru}$ is thickness of the Ru spacer, $M_\mathrm{s}$  is saturation magnetization, $J_\mathrm{bl}$ ($J_\mathrm{bq}$)is bilinear (biquadratic) interlayer coupling, $H_\mathrm{u}$ is uniaxial anisotropy field, $\gamma$ is gyromagnetic ratio, $A_\mathrm{ex}$ is exchange constant, $\alpha$ is dimensionless damping constant.}
    \begin{ruledtabular}
    \label{tab:Param}
    \begin{tabular*}{\textwidth}{@{\extracolsep{\fill}}l|ccccccccc}
         Parameter ~ & \makecell{ $t_\mathrm{CoFeB}$ \\ (nm) } & \makecell{$t_\mathrm{Ru}$ \\ (nm) } & \makecell{$M_\mathrm{s}$ \\ (kA/m) }& \makecell{$J_\mathrm{bl}$ \\ $(\mathrm{mJ}/\mathrm{m}^2)$} & \makecell{$J_\mathrm{bq}$ \\ $(\mathrm{mJ}/\mathrm{m}^2)$} &\makecell{ $H_\mathrm{u}$ \\ (kA/m)} & \makecell{ $\gamma/(2\pi)$ \\ (GHz/T)} & \makecell{ $A_\mathrm{ex}$ \\ (pJ/m) } & $\alpha$ \\  \hline
         Value \rule{0pt}{2.5ex}  & $15\pm 1 \ssymbol{1}$ & $0.6\pm0.2 \ssymbol{1}$ & $1299\pm14 \ssymbol{2}$ & $-0.67\pm0.02 \ssymbol{2}$ & $-0.28\pm0.02 \ssymbol{2}$ & $1.5\pm 0.2^\ddagger$ & $30^\mathsection$ & $15^\mathsection$ & $0.004^\mathsection$\\
    \end{tabular*}
    \end{ruledtabular}
    \begin{tabular*}{\textwidth}{l}
 $\ssymbol{1}$ Measured from TEM images. \\ $\ssymbol{2}$ Obtained from the simultaneous fit of OOP and IP hysteresis loops. \\ $^\ddagger$ Estimated from the offset of the out-of-plane polarized (acoustic) mode synthetic antiferromagnetic resonance. \\ $^\mathsection$ These parameters were fixed \cite{vanatka2021spin}.
\end{tabular*}
\end{table*}

\begin{figure}
\includegraphics{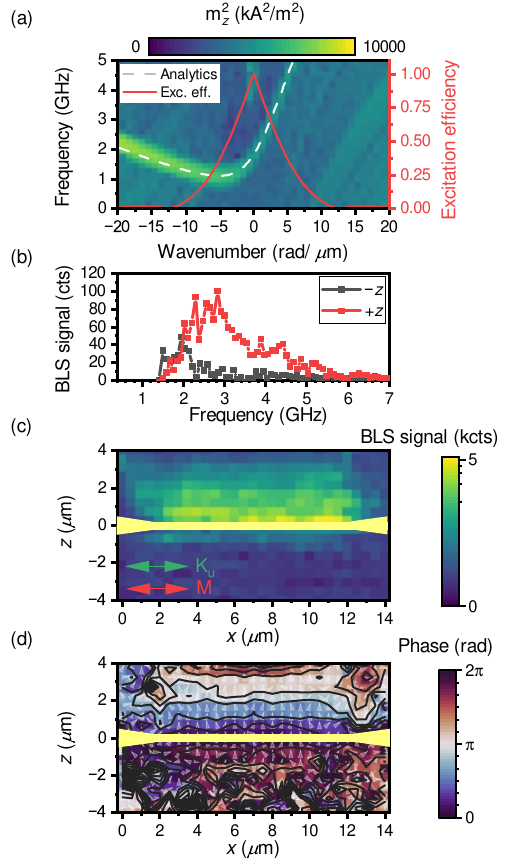}
\caption{\label{fig:uniDir} Unidirectional excitation of the spin waves in SAF. (a) Dispersion relation obtained by analytic model and MuMax3 simulation. The red curve shows the excitation efficiency of the used antenna. (b) BLS intensity measured approximately $1.5\,\mu$m away from the antenna in the positive (red symbols) and the negative (black symbols) $z$-direction. (c) two-dimensional mapping of the spin-wave intensity at the frequency of $2.5\,$GHz. Arrows depict directions of uniaxial anisotropy and magnetizations. (d) Extracted spin-wave phase from five BLS measurements at $2.5\,$GHz.}
\end{figure}

\begin{figure*}
\includegraphics{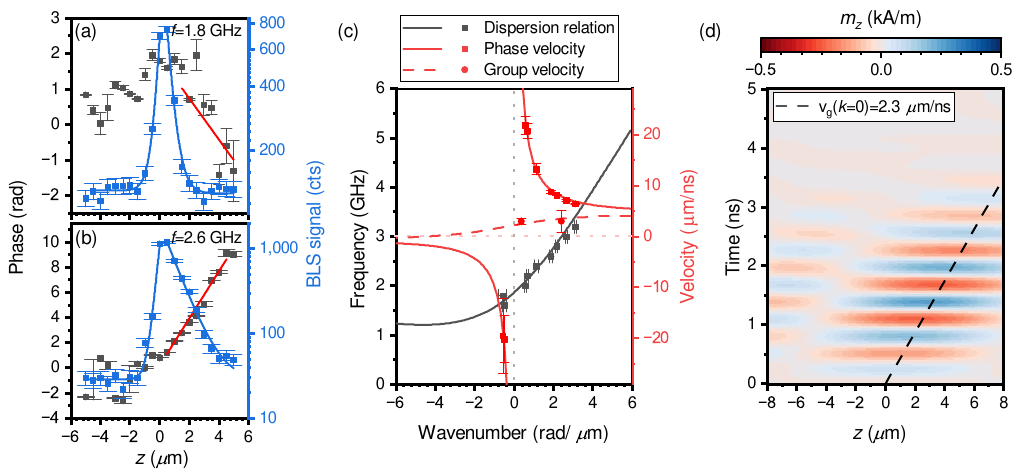}
\caption{\label{fig:Reversal} Phase velocity reversal upon $\Gamma$ point crossing. (a,b) Spin-wave intensity (blue squares) and spin-wave phase (black squares) at 1.8\,GHz (a) and 2.6\,GHz (b). The red and blue lines are fits of the linear evolution of phase and exponential decay of spin waves, respectively. (c) Dispersion relation and phase velocity obtained experimentally (black and red squares) and by analytic modeling (black and red solid lines) Group velocity measured by time-resolved BLS (red circles) and calculated from the analytical model (red dashed line). (d) Simulated space-time map of the zero-momentum spin-wave evolution. The configuration is the same as in Fig.~\ref{fig:Sketch}a. Black dashed line shows the group velocity. The appearance of the spin wave in the left part of the space-time map is caused by periodic boundary conditions used in the simulation.}
\end{figure*}

Out-of-plane polarized (acoustic) mode of SAF was measured by using vector network analyzer (VNA) in flip chip configuration \cite{adams2018critical} in two geometries: with the external magnetic field perpendicular and parallel to the easy axis of the SAF, see supplementary Fig.~S4. The positions of out-of-plane polarized (acoustic) mode, determined by fitting the Lorentz function for each applied external magnetic field, are shown in Fig.~\ref{fig:StaticChar}c. The experimental data agree well with the analytic model of out-of-plane polarized (acoustic) mode (shown in Fig.~\ref{fig:StaticChar}c as black and red lines) developed by Gallardo et al. \cite{gallardo2019reconfigurable, githubSWT}. This model is based on solving the following eigenproblem (assuming $k=0$): 
\begin{equation}
    \hat{A} \boldsymbol{m} = i \left( \frac{\omega }{\mu_0 \gamma} \boldsymbol{m} \right)
    \label{eq:eigenEq}
\end{equation}
where $\hat{A}$ is a system matrix, $\omega$ is a spin-wave frequency, $\gamma$ is a gyromagnetic ratio, and $\boldsymbol{m}$ is dynamic normalized magnetization vector in both layers. The system matrix depends on the sample geometry (CoFeB and Ru thicknesses), magnetization orientation of the individual CoFeB layers [obtained by energy minimization (see Eq.~\ref{eq:TotEnergy})], and on the magnetic properties (saturation magnetization $M_\mathrm{s}$, exchange constant $A_\mathrm{ex}$, uniaxial anisotropy field $H_\mathrm{u}$ and on interlayer exchange coupling $J_\mathrm{bl}$ and $J_\mathrm{bq}$). To calculate results shown in Figs.~\ref{fig:StaticChar}c,~\ref{fig:uniDir}a, and~\ref{fig:Reversal}c, we used parameters from Table~\ref{tab:Param}.

To obtain SAF dispersion, we can use the same analytic model (Eq.~\ref{eq:eigenEq}), which can be extended for $k \in \left[-20\,\mathrm{rad}/\mu\mathrm{m},20\,\mathrm{rad}/\mu\mathrm{m}\right]$ (gray line in Fig.~\ref{fig:uniDir}a). To further confirm the results of the analytic model we performed micromagnetic simulations in MuMax3 (see colormap in Fig.~\ref{fig:uniDir}a). The simulation cell size was set to ($10\times10$)\,nm$^2$ in the IP direction and 15\,nm in the OOP direction, i.e., we neglected the Ru spacer and assumed only one cell per one CoFeB layer. The simulation size was $2048\times2048\times2$ cells. The dispersions were calculated using the approach described in \cite{wojewoda2023observing}.

Before BLS experiments, we stabilized the single-domain magnetization state in wide-field Kerr microscope. The Neél vector pointed parallel to the stripline antenna (connected to RF generator), which was used to excite spin waves in the SAF layer. As the first BLS experiment, we swept the excitation frequency and compared the BLS spectra acquired 
at a distance of $1.5\,\mu\mathrm{m}$ from the antenna on both sides, see Fig.~\ref{fig:uniDir}b. Higher excitation efficiency on the positive side of the antenna is clearly visible. On the negative side, the spin wave signal above 2\,GHz is completely suppressed. At this region, the antenna cannot efficiently excite spin waves due to the small excitation efficiency of the antenna in the range of available wavenumbers, see Fig.~\ref{fig:uniDir}a, red line. The excitation efficiency was calculated using fast Fourier transform of the in-plane component of the field calculated by FEMM solver \cite{Meeker}. The signal below 2\,GHz visible on the negative side of the antenna originates from the dispersion region with negative group velocity ($k<-5\,\mathrm{rad}/\mu\mathrm{m}$). Note, that this signal can be entirely suppressed by modifying the excitation efficiency, i.e. by using a slightly wider antenna, thus the spin wave emission can be completely unidirectional. 

To further investigate this non-reciprocal behavior, we performed a 2-dimensional mapping of BLS signal at the frequency of 2.5\,GHz, shown in Fig.~\ref{fig:uniDir}c. At this frequency, we can observe unidirectional excitation of uniform spin waves. We have performed full phase-resolved mapping of spin waves with use of five different measurements \cite{wojewoda2023phase}. The results (Fig.~\ref{fig:uniDir}d) show clear spatially uniform evolution of the phase in the direction perpendicular to the antenna in the positive $z$-direction. On the other side of the antenna, no phase evolution was observed.

As the next experiment, we performed two full phase-resolved linescans in the direction perpendicular to the antenna at the frequencies of 1.8\,GHz and 2.6\,GHz, and extracted the intensity and phase of spin waves (Fig.~\ref{fig:Reversal}a, b). For 1.8\,GHz, we can observe a negative phase evolution in the positive values of $z$ (upper side of the antenna in Fig.~\ref{fig:uniDir}c,d). On the other hand, at the frequency of 2.6\,GHz, we can observe a positive phase evolution on the same side of the antenna. The intensity data show prominent non-reciprocity, where exponential decay is visible only towards the positive direction [the decay constants for negative direction are $0.65\pm0.04\,\mu m$ (1.8\,GHz), $0.49\pm0.03\,\mu m$ (2.6\,GHz), and for positive direction $0.84\pm0.06\,\mu m$ (1.8\,GHz), $1.9\pm0.1\,\mu m$ (2.6\,GHz)]. By repeating this experiment at different frequencies we obtained full dispersion relation (gray squares) and frequency-dependent phase velocity (red squares), see Fig.~\ref{fig:Reversal}c. We also performed time-resolved BLS experiments and extracted group velocities (see red circles in Fig.~\ref{fig:Reversal}c, and supplementary Fig.~S5). The experimental data agree well with our analytical model (see solid and dashed lines in Fig.~\ref{fig:Reversal}c).

Spin waves at the $\Gamma$ point ($k=0$) are shifted up by approx. 1.9\,GHz, which agrees well with the data in Fig.~\ref{fig:StaticChar}c, and is caused by the uniaxial anisotropy. The anisotropy and the dipolar interaction induce the monotonous and smooth dispersion through the $\Gamma$ point and leads to unidirectional group velocity (see dashed line in Fig.~\ref{fig:Reversal}c). The group velocity is conserved despite the change of the wavenumber (phase velocity) from positive to negative. 

This property leads to interesting phenomena, where the wave with zero momentum ($k=0$), can still propagate in space. To demonstrate this, we performed micromagnetic simulation where the dynamics was excited by OOP sinc pulse centered at $z=0$\,$\mu$m. The evolution of the whole excited spin wave packet in space and time is shown in supplementary Fig.~S6. If we filter only wavevectors with $k\to0$ we can clearly see the propagation of this "wave" towards the positive value of $z$ (see Fig.~\ref{fig:Reversal}d). The phase velocity $v_\mathrm{p}\to\infty$, while the group velocity $v_\mathrm{g}=2.3\,\mu \mathrm{m}/\mathrm{ns}$ is given by the dispersion relation (see Fig.~\ref{fig:Reversal}c), which is also in agreement with simple formula given in \cite{millo2023unidirectionality} ($v_\mathrm{g}=2.31\pm0.03\,\mu \mathrm{m}/\mathrm{ns}$). The orientation of the unidirectional propagation can be reversed by switching the SAF magnetization (see supplementary Fig.~S6).


In conclusion, we observed unidirectional propagation of zero-momentum magnons in synthetic antiferromagnet. Inherent non-reciprocity of spin waves in synthetic antiferromagnets with uniaxial anisotropy shapes the dispersion relation into a smooth and monotonous function in the vicinity of the $\Gamma$ point. This situation leads to unidirectional propagation of spin waves with negative, zero and positive wavenumbers. This phenomenon was observed by intensity-, phase-, and time-resolved BLS microscopy and further corroborated by micromagnetic simulations and analytical modelling. Such unusual dispersion relation, up to our knowledge, only occurs for planetary waves (e.g. Yanai waves), where the control of experimental condition is impossible \cite{Khouider2013}. Therefore, our system offers many possibilities for further investigations of related phenomena e.g., propagation of wavepackets with bidirectional phase velocities, or spin wave accumulation on domain boundaries. In practical applications such unidirectional system can pave the way towards designing new non-reciprocal elements for signal processing and unconventional spin-wave computing.

\begin{acknowledgements}
The authors thank to Roman Verba for discussion on analytical models for SAFs and Radek Kalousek for discussion on consequences of smooth and monotonous dispersion relation through $\Gamma$-point. This research was supported by the Horizon Europe EIC Pathfinder program, project ID 101098651 (METASPIN) and by the Grant Agency of the Czech Republic, project no. 23-04120L. CzechNanoLab project LM2023051 is acknowledged for the financial support of the measurements and sample fabrication at CEITEC Nano Research Infrastructure. Specific research projects IDs CEITEC VUT-J-23-8435 and CEITEC VUT-J-23-8411 are also acknowledged. O.W. was supported by Brno PhD talent scholarship. A.V.C. acknowledges support by the Austrian Science Fund (FWF) under Grant No. I 4917-N (MagFunc).
\end{acknowledgements}
\bibliography{apssamp}

\pagebreak
\clearpage
\widetext
\title{Supplemental Material: Unidirectional propagation of zero-momentum magnons}
\maketitleSupp
\renewcommand\figurename{FIG. S}
\onecolumngrid

\vspace{50pt}
\begin{figure*}[!b]
\includegraphics{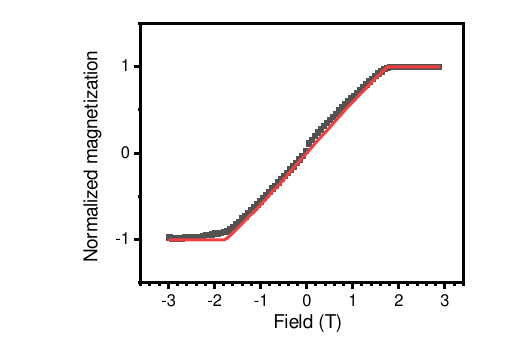}
\caption{\label{fig:OOPVSM} Out-of-plane hysteresis loop of the synthetic antiferromagnet sample measured by vibrating sample magnetometry. The data are fitted by minimizing Eq.~1}
\end{figure*}

\begin{figure*}
\includegraphics{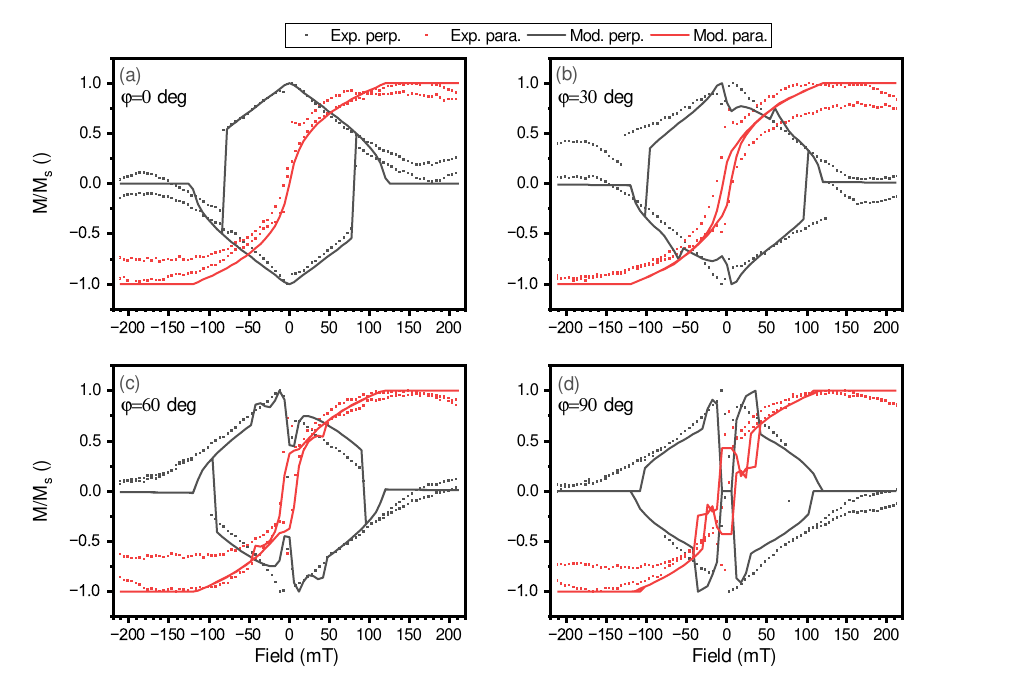}
\caption{\label{fig:Kerr} In-plane hysteresis loops where the angle between the external field and the uniaxial anisotropy was 0 deg (a), 30 deg (b), 60 deg (c), and 90 deg (d). The solid lines are calculated by minimizing Eq.~1, and the squares represent the experimental data measured by a wide-field Kerr microscope. Red (black) dots show magnetization projection to the parallel (perpendicular) to the external field direction.}
\end{figure*}

\begin{figure*}
\includegraphics{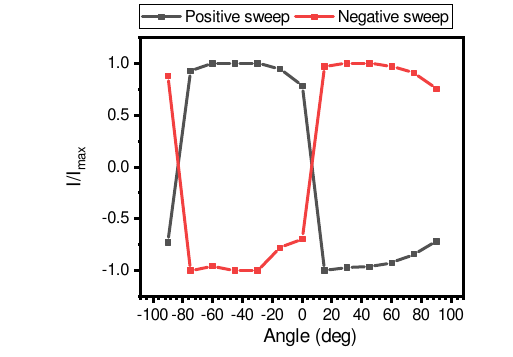}
\caption{\label{fig:KerrSwitch} Wide-field Kerr microscopy measurement of the remanent component of the magnetization vector, pointing parallel to the direction of the uniaxial anisotropy. The sample was measured at different angles of the external magnetic field and with different field sweep directions.}
\end{figure*}

\begin{figure*}
\includegraphics{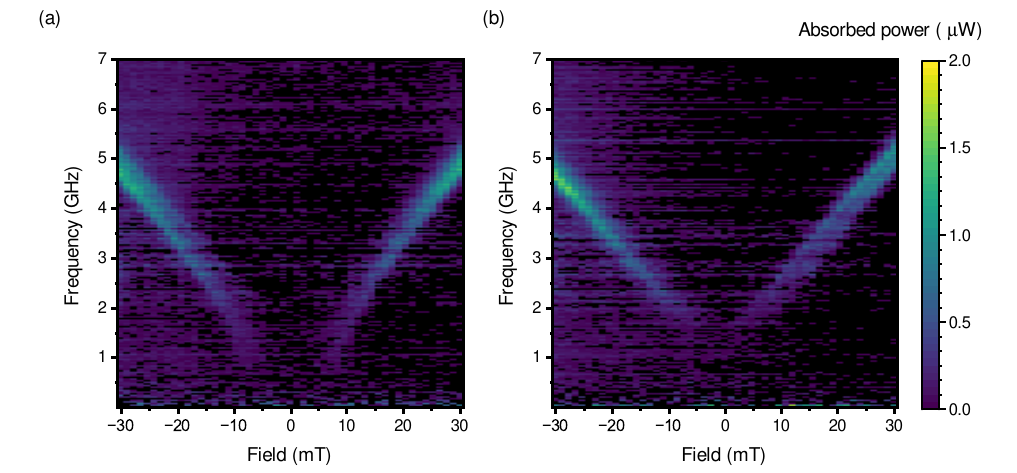}
\caption{\label{fig:SAFMR} Out-of-plane polarized (acoustic) mode of SAF measured on coplanar waveguide in flip chip configuration with the excitation field perpendicular to external bias field. The total absorbed power calculated from S12 parameter measured by vector network analyzer. The uniaxial anisotropy was oriented perpendicularly (a), and parallel~(b) to the external magnetic field direction.}
\end{figure*}

\begin{figure*}
\includegraphics{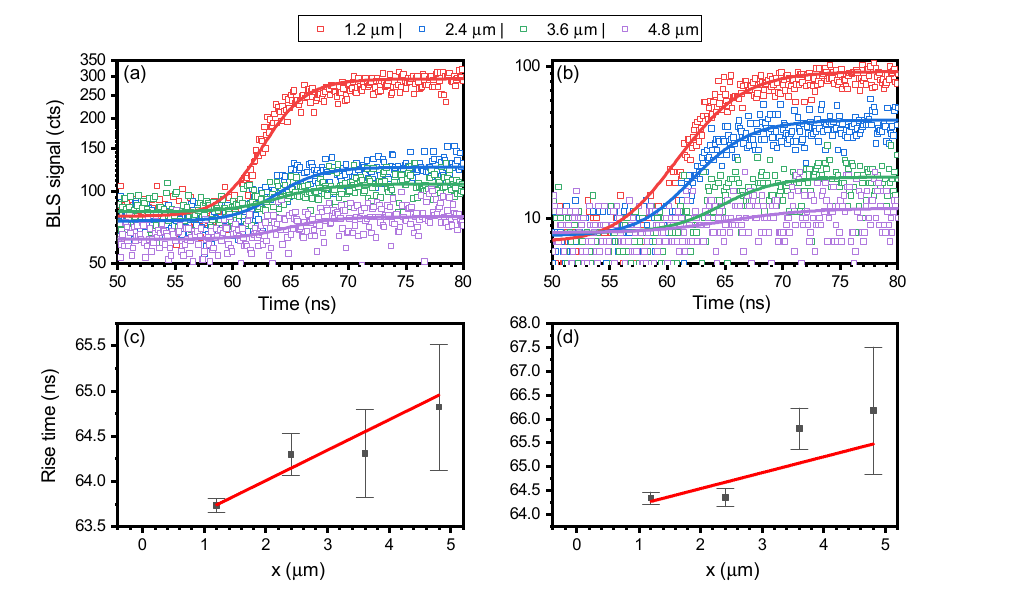}
\caption{\label{fig:TRBLS} (a,b) Time-resolved BLS signal acquired at different distances from the exciation antenna at the frequencies of 2\,GHz (a) and 3\,GHz (b). The rise times were calculated by fitting the rising edge by a logistic function. Panel (c) shows the data measured at the frequency of 2\,GHz and panel (d) shows the data measured at the frequency of 3\,GHz. The data are subsequently fitted with linear function with resulting group velocities of $3.0\pm0.6\,\mu$m (c) and $3\pm2\,\mu$m (d).}
\end{figure*}

\begin{figure*}
\includegraphics{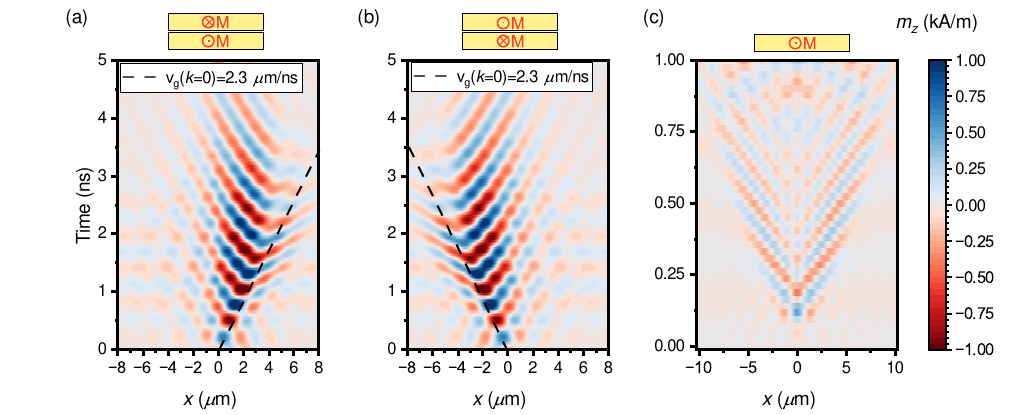}
\caption{\label{fig:tvsz} (a, b) The spatio-temporal map of the sinc pulse (both in time and space) excitation for two opposite configurations of SAF. We can observe that the zero-momentum ($k=0$) spin waves propagate to the right (a) or left (b) with the group velocity of $\left|2.3\right|\,\mu\mathrm{m}/\mathrm{ns}$ (black dashed lines). The waves with lower (higher) group velocities and negative (positive) phase velocities, are above (below) the dashed lines. The change in the phase velocity direction (the sign of wavenumber), exhibits itself in this representation as the change of the slope of the wavefronts. (c) For comparison, we present also a simulation of a 30\,nm thick ferromagnetic slab with the same material parameters (see Table~1 in the main text).}
\end{figure*}

\end{document}